\documentclass[twocolumn,showpacs,superscriptaddress,prb,amsmath,amssymb,letterpaper]{revtex4}
\usepackage{times}
\usepackage{amsmath,bm,amsfonts}
\usepackage{dcolumn}
\usepackage{graphicx}
\usepackage{latexsym}

\renewcommand{\v}[1]{\ensuremath{\mathbf{#1}}} 

\newcommand{\avg}[1]{\left< #1 \right>} 
\newcommand{\pd}[2]{\frac{\partial #1}{\partial #2}}
\newcommand{\ket}[1]{\left| #1 \right>} 
\let\baraccent=\= 
\renewcommand{\=}[1]{\stackrel{#1}{=}} 
\newcommand{\Dk}{\frac{\mathrm{d}^3k}{(2\pi)^3}}
\newcommand{\bk}{\v{k}}

\newcommand{\bq}{\v{q}}

\newcommand{\bQ}{\v{Q}}
\newcommand{\bS}{\v{S}}

\newcommand{\tu}{\tilde{u}}
\newcommand{\tv}{\tilde{v}}
\newcommand{\HM}{\mathcal{H}}
\newcommand{\ekt}{\tilde{\epsilon}_{\bk}}
\newcommand{\ncr}{n_{\textrm{cr}}}
\newcommand{\meff}{\tilde{\mu}}
\newcommand{\sbar}{\bar{s}}
\newcommand{\sgn}[1]{\mathrm{sgn}{#1}}
\newcommand{\eref}[1]{(\ref{#1})}

\begin{document}

\title{Theory of Magnetic Field-Induced Bose-Einstein Condensation of Triplons in
Ba$_{3}$Cr$_{2}$O$_{8}$}

\author{Tyler \surname{Dodds}}
\affiliation{Department of Physics, University of Toronto,
Toronto, Ontario M5S 1A7, Canada}

\author{Bohm-Jung \surname{Yang}}
\affiliation{Department of Physics, University of Toronto,
Toronto, Ontario M5S 1A7, Canada}

\author{Yong Baek \surname{Kim}}
\affiliation{Department of Physics, University of Toronto,
Toronto, Ontario M5S 1A7, Canada}
\affiliation{School of Physics,
Korea Institute for Advanced Study, Seoul 130-722, Korea}

\date{\today}

\begin{abstract}
Motivated by recent experiments on Ba$_{3}$Cr$_{2}$O$_8$, a new
spin-dimer compound with spin-1/2 moments of Cr$^{5+}$ ions,
we theoretically investigate the field-induced magnetic ordering
in this material in view of the Bose-Einstein condensation (BEC) of
triplet excitations (triplons).
We apply the self-consistent Hartree-Fock-Popov (HFP) approach
to a microscopic Hamiltonian, using
the realistic triplon dispersion measured in an inelastic neutron
scattering experiment.
In particular, we ask to what extent the BEC of dilute triplons
near the critical field can explain
the magnetic ordering in this material. For example,
we investigate the temperature range where the BEC picture of
triplons can be applied via the HFP approach. We also determine the
temperature regime where a quadratic approximation of
the triplon dispersion works. It is found that the strength of
the effective repulsive interaction between triplons is
much weaker in Ba$_{3}$Cr$_{2}$O$_8$ than in the
canonical spin-dimer compound TlCuCl$_{3}$.
Small effective repulsive interaction in combination
with the narrow band of triplons leads to higher
density of triplons $n_{\rm cr}$ at the critical point.
It turns out that the combined effect points to a bigger HFP
correction $U n_{\rm cr}$ in Ba$_3$Cr$_2$O$_8$ than in TlCuCl$_{3}$.
Nonetheless, the HFP approach provides a reasonable explanation of
the transverse
magnetization and the specific heat data
of Ba$_{3}$Cr$_{2}$O$_8$.
\end{abstract}


\maketitle

\section{\label{sec:intro} Introduction}

Magnetic-field-induced quantum phase transitions in spin dimer
systems have provided
excellent playgrounds for the investigation of novel universality
classes of zero temperature
quantum phase transitions.\cite{GiamarchiReview}
These systems possess non-magnetic spin singlet ground states
with a spin gap to
triplet excitations (triplons). When a magnetic field $H$ is applied,
a quantum phase
transition occurs at a critical field $H_{c}$, where the spin gap
closes and the lowest
triplet excitation condenses. At $H > H_{c}$, the average triplon
density is finite
and can be controlled by the applied magnetic field.

The resulting ground states are determined by a delicate balance
between the
kinetic energy and the repulsive interaction between triplons.\cite{Rice}
On one hand, if the triplon hopping processes are suppressed by
frustration or
the repulsive interaction dominates, the condensed triplons
may form a superlattice
with broken translational symmetry, leading to magnetization plateaus.
This is known to occur, for instance,
in SrCu$_2$(BO$_3$).\cite{Kodama, kageyama99}
On the other hand, if the magnetic interaction does
not have much frustration
or the kinetic energy dominates, the ground state
can be described as a Bose-Einstein
condensate (BEC) of triplons and form a homogeneous
magnetically ordered state.
In this case, the magnetically ordered state at $H > H_{c}$
supports a staggered
magnetization transverse to the field direction, leading to
a canted antiferromagnetic state
(until the system eventually becomes fully polarized
as $H$ increases).
This type of behavior has been observed, for example,
in three-dimensionally
coupled spin dimer systems TlCuCl$_3$\cite{ruegg03, oosawa}
and BaCuSi$_2$O$_6$,\cite{jaime04} 
which exhibits
unconventional critical behavior.\cite{sebastian06, Ruegg2, Kramer2, Laflorencie} Furthermore,
recent discoveries of A$_3$M$_2$O$_8$,\cite{kofudispersion}
where A = Ba or Sr, and M = Cr or Mn, have provided a lot of
excitement for spin dimer system research,
as these systems may represent a variety
of different spin dimer
interactions and quantum ground states.

\begin{figure}[t]
\includegraphics[width = 6.5 cm]{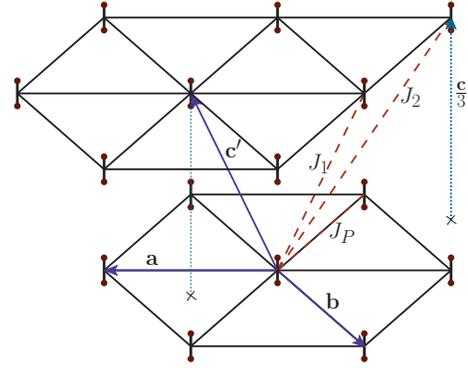}
\caption{(colour online)
Schematic diagram showing two neighboring triangular
lattice planes of dimers in Ba$_3$Cr$_2$O$_8$.
Two primitive lattice vectors $\v{a}$ and $\v{b}$ are
shown in the lower plane.
The third primitive lattice vector $\v{c}'$ connects central dimers
in the neighboring planes.
We set the vertical distance between neighboring planes by $\v{c}$/3.
Here we use $J_P$ to indicate the inplane
nearest-neighbour interdimer coupling. $J_1$ ($J_2$) denotes
the nearest-neighbour (next nearest-neighbor)
inter-plane dimer coupling.}
\label{fig:twoplanes}
\end{figure}

In this work, we present a theory of the magnetic field-induced
quantum phase transition
discovered in Ba$_3$Cr$_2$O$_8$, where Cr$^{5+}$ carries an $S$=1/2
moment (3d$^1$).\cite{kofutransition}
Low temperature bulk susceptibility shows that this compound
does not have any magnetic long-range order down to 1.5 K in
the absense of an
external magnetic field.\cite{nakajima,singh}
When the external magentic field $H$ reaches $H_{c1} \sim$ 12 T, a
field-induced transition to
a magnetically ordered state occurs and a fully polarized state
arises at $H > H_{c2} \sim$ 23 T.\cite{kofutransition}
In this compound, two neighboring $S$=1/2 Cr$^{5+}$ ions lying
along the $\textbf{c}$
direction form a singlet dimer.
In the $ab$-plane, these dimer singlets are coupled into
triangular lattices, which
are stacked along the $\textbf{c}$ direction
(see Fig.\ref{fig:twoplanes}).
According to recent elastic and inelastic neutron scattering
measurements,\cite{kofudispersion}
Ba$_3$Cr$_2$O$_8$ is an excellent model system for
weakly coupled
$S=$1/2 quantum spin dimers, featuring strong intradimer coupling
of $J_{0}$=2.38(2) meV
and weak interdimer couplings less than
0.52(2) meV.\cite{kofudispersion}
Because of the orbital degeneracy of the Cr$^{5+}$
ion, there is a structural
transition around 70 K via a Jahn-Teller distortion,
relieving the frustration.
As a consequence, spatially anisotropic
interdimer couplings arise.
The relative orientations of the anisotropic interdimer
couplings are
described in Fig.\ref{fig:anisotintns}.
It was confirmed that the magnetically
ordered state has a commensurate and collinear transverse
spin component for $H_{c1} < H < H_{c2}$.\cite{kofutransition}
This is in contrast to the case of
Ba$_3$Mn$_2$O$_8$ with orbitally
non-degenerate $S$=1 Mn$^{5+}$ ions (3d$^2$),
where the geometric frustration
and single-ion
anisotropy lead to incommensurate spiral order upon
triplon condensation.\cite{Ba3Mn2O8}

We first consider the Heisenberg spin Hamiltonian using the
spin exchange couplings
determined by inelastic neutron scattering measurements.
\cite{kofudispersion}
Applying the bond operator formalism, we obtain the dispersion
of the lowest energy triplet excitations.\cite{matsumotoprb,matsumotoprl}
We confirm that
the Hamiltonian written in terms of bond operators at the
quadratic level,
neglecting singlet fluctuation, leads to the same triplon dispersion
as determined in experiment in Ref.\onlinecite{kofudispersion}.

We then use the Hartree-Fock-Popov (HFP) approximation combined
with the
realistic triplon dispersion to interpret two different experimental
data sets, those of
M. Kofu \textit{et al.} in Ref.\onlinecite{kofutransition}
and A. A. Aczel \textit{et al.} in Ref.\onlinecite{aczeldata}.
In particular, we would like to understand to what extent
Ba$_3$Cr$_2$O$_8$ is a good candidate for
the BEC of triplons in comparison to other three-dimensionally
coupled spin-dimer systems such as
TlCuCl$_3$.\cite{nikuni,misguich,yamada08}
Within the HFP analysis, we determine the effective inter-triplet
repulsion $U$ and the zero-field
spin gap $\Delta$ in Ba$_3$Cr$_2$O$_8$. It turns out that
the strength of the effective repulsive interaction $U$ between
triplons in Ba$_3$Cr$_2$O$_8$
is an order of magnitude smaller than that of TlCuCl$_{3}$,
\cite{misguich} and
smaller than the bandwidth of the triplons as well. This suggests
that the system is
indeed in the regime where the kinetic energy dominates.
In addition, the shape of the dispersion near the triplon band minimum
results in the large effective mass of triplons in
Ba$_3$Cr$_2$O$_8$.
As a result, the relation $[H_c(T)-H_c(0)] \propto T^{3/2}$ in
three dimensions for quadratic triplon dispersion works only
at $T < 0.1$ K,
while it works at $T < 1$ K in TlCuCl$_3$.
The HFP approach is used to describe other
experimental measurements such as the
longitudinal and transverse staggered
magnetizations and the heat capacity.
Despite the simplicity of the theoretical
approach,
the HFP approximation is found to explain these physical
properties even quantitatively.

The rest of the paper is organized as follows.
In Sec.\ref{sec:bondop} we use the bond operator
approach to obtain the triplet dispersion
from the microscopic Hamiltonian. Theoretical description
of the triplon Bose-Einstein condensation within the HFP
approximation is discussed in detail in Sec.\ref{sec:hfp}.
In Sec.\ref{sec:phasediagram}, we apply the HFP approach
to explain the experimental data and draw the phase diagram
in the plane of $H_{c}$ versus T. The HFP approach
is applied to describe the heat capacity in Sec.\ref{sec:specheat}
and the magnetization measurements in Sec.\ref{sec:magnetization}.
Finally, in Sec.\ref{sec:discussion}, we summarize
our result, and discuss possible limitations and
extensions of the current work.

\section{Triplon Dispersion via Bond-Operator Approach}
\label{sec:bondop}

Since the intradimer exchange interaction dominates
all the other interdimer couplings in Ba$_3$Cr$_2$O$_8$,
it is natural to take advantage of
the bond-operator representation of
the singlet and triplet dimer states.\cite{chubukov,sachdevbhatt}
This is achieved by
placing one singlet or triplet boson on each dimer, to represent the
states
\begin{align}\label{eqn:bondopstates}
&\ket{s} = s^{\dag}\ket{0} =
\frac{\ket{\uparrow\downarrow}-\ket{\downarrow\uparrow}}{\sqrt{2}},
\nonumber\\
&\ket{t_0} = t_0^{\dag}\ket{0} =
\frac{\ket{\uparrow\downarrow}+\ket{\downarrow\uparrow}}{\sqrt{2}},
\nonumber\\
&\ket{t_+} = t_+^{\dag}\ket{0} =
-\ket{\uparrow\uparrow},
\nonumber\\
&\ket{t_-} = t_-^{\dag}\ket{0} =
\ket{\downarrow\downarrow},
\end{align}
where the quantization $z$-axis is taken to be the applied field direction.
The triplet states $\ket{t_m}$ ($m=1,0,-1$) are chosen
as the $S_z$ eigenstates
satisfying $S_z$$\ket{t_m}$=$m\hbar$$\ket{t_m}$.
The hard-core constraint
$s^{\dag}s+t^{\dag}_0t_0+t^{\dag}_+t_++t^{\dag}_-t_-=1$
is enforced on each dimer, ensuring that the physical state is
exactly one of the four above.

The two spin operators constituting a dimer
can be rewritten in terms of the bosonic bond operators as
\begin{align}
S_{1\alpha} &= \frac{1}{2}\left(s^{\dag}t_{\alpha} + t^{\dag}_{\alpha}s
-i\epsilon_{\alpha\beta\gamma}t^{\dag}_{\beta}t_{\gamma}\right),
\nonumber \\
S_{2\alpha} &=
\frac{1}{2}
\left(-s^{\dag}t_{\alpha} - t^{\dag}_{\alpha}s
-i\epsilon_{\alpha\beta\gamma}t^{\dag}_{\beta}t_{\gamma}\right)
\end{align}
with $\alpha\in\{x,y,z\}$
and $\epsilon$ the totally antisymmetric tensor.
This form gives the correct matrix elements in the
singlet-triplet Hilbert space.
We define $\ket{t_{\alpha}}$ triplons as eigenstates with
$S_{\alpha}\ket{t_{\alpha}}=0$
so that $t_z=t_0$, $t_x = \frac{1}{\sqrt{2}}\left(t_-+t_+\right)$
and $t_y = \frac{i}{\sqrt{2}}\left(t_--t_+\right)$.
We henceforth assume a sum over
repeated indices.

\begin{figure}[htp]
\includegraphics[width = 6.5 cm]{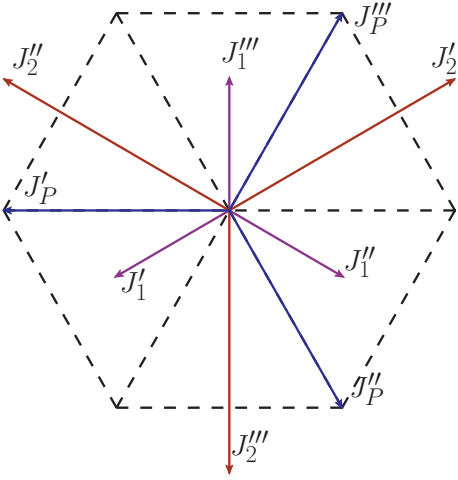}
\caption{(colour online)
In-plane projection of the strengths and directions of the
different anisotropic interdimer couplings considered in this work.
The couplings $J_P$ are in-plane nearest neighbor couplings,
which are represented by
dashed lines. $J_1$ ($J_2$) is the first (second)
nearest-neighbour interplane coupling.}
\label{fig:anisotintns}
\end{figure}

Besides connecting spin operators to singlet and triplet
boson operators,
the bond operator formalism
naturally yields a
triplon dispersion from the microscopic Hamiltonian.
We consider a Heisenberg spin Hamiltonian
with intra-dimer coupling, and coupling between nearby dimers.
We include nearest neighbor interactions between
dimers on the same plane. Between
adjacent planes, we include interactions
between first and second-nearest neighboring dimers.
The tetrahedrally-coordinated $3d^1$ electron in
Cr$^{5+}$ has $e_g$ orbital degeneracy and
undergoes Jahn-Teller distortion. This gives rise to spatially
anisotropic interdimer interactions.\cite{kofudispersion}
The inplane projection of the anisotropic interdimer
interactions are described in Fig.\ref{fig:anisotintns}.

The intra-dimer coupling has a strength of $J_0=2.38$ meV.
Table \ref{tab:anisotropicinteractions}
displays the values and directions
of the couplings to nearby dimers depicted in detail in
Fig.\ref{fig:anisotintns}.
These strengths have been determined in Ref.\onlinecite{kofudispersion}
by
fitting an RPA dispersion to the
triplon dispersion measured by inelastic
neutron scattering.

\begin{table}
\begin{tabular}{|c| c| c| c|}
\hline \hline
$m$ & $J_m$ (meV) & $\Delta\v{r}_m$ \\
\hline \hline
1 & $J_P' = 0.1$ & $\v{a}$ \\
2 & $J_P'' = 0.07$ & $\v{b}$ \\
3 & $J_P''' = -0.52$ & $-\v{a}-\v{b}$ \\
4 & $J_1' = 0.08$ & $2\v{a}/3 +\v{b}/3 + \v{c}/3$  \\
5 & $J_1'' = -0.15$ & $-\v{a}/3 +\v{b}/3 + \v{c}/3$ \\
6 & $J_1''' = 0.1$ & $-\v{a}/3 -2\v{b}/3 + \v{c}/3$ \\
7 & $J_2' = 0.04$ & $-4\v{a}/3 -2\v{b}/3 + \v{c}/3$ \\
8 & $J_2'' = 0.1$ & $2\v{a}/3 -2\v{b}/3 + \v{c}/3$ \\
9 & $J_2''' = 0.09$ & $2\v{a}/3 +4\v{b}/3 + \v{c}/3$\\
\hline \hline
\end{tabular}
\caption{Interaction strength $J_m$ and relative distance
$\Delta\v{r}_m$ for the nine different anisotropic interdimer
couplings as described in Fig.\ref{fig:anisotintns}.}
\label{tab:anisotropicinteractions}
\end{table}
There are four possible spin-spin interactions between two dimers
at sites $i$ and $j$.
However, at quadratic order in the $t$ bosons,
$\v{S}_{i1}\cdot\v{S}_{j1}= \v{S}_{i2}\cdot\v{S}_{j2} =
-\v{S}_{i1}\cdot\v{S}_{j2} = -\v{S}_{i2}\cdot\v{S}_{j1}$.
All of the potential spin-spin Heisenberg interactions between
two dimers thus reduce to a single effective interaction $J_m$.
With this, we take the Hamiltonian to be
\begin{align}
\HM = \sum_i J_0\bS_{i1}\cdot\bS_{i2} +
\sum_i\sum_{m=1}^{9}
J_m\bS_{i,1}\cdot\bS_{i+\delta{\v{r}_m,1}},
\label{eqn:hamiltonianrealspace}
\end{align}
where the intra-dimer interaction can be written as
\begin{align}
J_0\v{S}_{1i}\cdot\v{S}_{2i} = J_0\left(-\frac{3}{4}s^{\dag}_is_i
+\frac{1}{4}t^{\dag}_{i\alpha}t_{i\alpha}\right)
.\end{align}
We now investigate the ground state and the excitations of this
Hamiltonian in the low temperature regime where
the intra-dimer interaction dominates.
Singlets are energetically favourable in this case,
with macroscopic occupation, we can take them to be condensed.
This process ignores singlet fluctuations by replacing the
spin operators $s^{\dag}_i$ and $s_i$ with
a complex number $\sbar$, creating a uniform condensate.

Condensing the singlets leaves a Hamiltonian describing the triplets.
Inter-dimer interactions lead to triplon excitation dispersion
in momentum
space. Our analysis considers the low triplon density limit, so we
retain terms up to quadratic order in the $t$ operators.
To impose the hardcore constraint we introduce a site-dependent chemical potential $\mu_i$.
This adds to the Hamiltonian the constraint term
\begin{align}
\sum_i\mu_{i}\left(
1-s^{\dag}_is_i-t^{\dag}_{i\alpha}t_{i\alpha}\right)
.\end{align}
To make a tractible analysis, the above constraint is
enforced in a mean-field manner taking $\mu_i=\mu$, on average over the entire lattice.
In the momentum representation, this
gives $\sbar^2+\int\Dk t^{\dag}_{\bk\alpha}t_{\bk\alpha}=1$.
At $T=0$, the values of $\mu$ and $\sbar$ are
determined from the saddle point
condition. First, $\avg{\pd{\HM}{\mu}}=0$ enforces the constraint
$0=\sum_i\left(
1-\avg{s^{\dag}_is_i}-\avg{t^{\dag}_{i\alpha}t_{i\alpha}}\right)$, as
expected.
Second, the condition $\avg{\pd{\HM}{\sbar}}=0$ minimizes the
ground-state energy with respect to the singlet density.

Our Hamiltonian obtains a quadratic form in the momentum space:
$\HM = N_d\epsilon_0 + \HM_0 + \HM_\pm$.
Here $N_d$ denotes the number of dimers on the lattice.
The $t_0$ triplons interact with themselves but not
the other triplon
species. They contribute with the quadratic Hamiltonian
\begin{align}
\HM_0 =
\frac{1}{2}\sum_{\bk}
\begin{pmatrix}
t^{\dag}_{\bk0}&
t_{-\bk0}
\end{pmatrix}
\begin{pmatrix}
A_{\bk} & B_{\bk}\\
B^*_{\bk} & A^*_{\bk}
\end{pmatrix}
\begin{pmatrix}
t_{\bk0} \\ t^{\dag}_{-\bk0} \end{pmatrix}
,\end{align}
where $A_{\bk} = \frac{J_0}{4} - \mu + B_{\bk}$, with
\begin{align}
B_{\bk} = -\sbar^2
\sum_mJ_m\cos{(\bk\cdot\Delta\v{r}_m)}.
\end{align}
$\HM_0$ can be diagonalized by the Bogolibov transormation
$\gamma_{\bk0} = u_{\bk}t_{\bk0} +
v_{\bk}t^{\dag}_{-\bk0}$, with quasiparticle energy\cite{blaizotripka}
\begin{align}
\omega_{\bk}=\sqrt{A^2_{\bk}-B^2_{\bk}}=
\sqrt{(\frac{J_0}{4}-\mu)^2+2(\frac{J_0}{4}-\mu)B_{\bk}}.
\label{eqn:bondopqpenergy}
\end{align}
Neither $t_0$ nor $\gamma_0$ are subject to Zeeman splitting
by the external field.
However, the $t_+$ and $t_-$ triplons are split. Furthermore, they
interact with each other. The resultant quadratic Hamiltonian ${\cal H}_{\pm}$ is
\begin{align}
&\HM_\pm =
\frac{1}{2}\sum_{\bk}
\Psi^{\dag}_\bk
\begin{pmatrix}
A_{\bk}-h&0&0&B_{\bk} \\
0&A_{\bk}+h&B_{\bk}&0 \\
0&B_{\bk}&A_{\bk}-h&0 \\
B_{\bk}&0&0&A_{\bk}+h
\end{pmatrix}
\Psi_\bk
\label{eqn:hamiltoniankspace}
,\end{align}
where $h = g\mu_BH$ and $\Psi^{\dag}_\bk$=($t^{\dag}_{\bk+}$ $t^{\dag}_{\bk-}$ $t_{-\bk+}$ $t_{-\bk-}$).
$\HM_\pm$ is
diagonalized into quasiparticles $\gamma_{\bk\pm}$
with energy $\omega_{\bk}\mp h$, where
\begin{align}
\gamma^{\dag}_{\bk+} = u_{\bk+}t_{\bk+}^{\dag} + v_{\bk+}t_{-\bk-}
,\nonumber \\
\gamma^{\dag}_{\bk-} = u_{\bk-}t_{\bk-}^{\dag} + v_{\bk-}t_{-\bk+},
\end{align}
\begin{align}
u_{\bk+}&=\frac{B_{\bk}}{\sqrt{2\omega_{\bk}(A_{\bk}-\omega_{\bk})}},~
v_{\bk+}=\frac{A_{\bk}-\omega_{\bk}}{\sqrt{2\omega_{\bk}(A_{\bk}-\omega_{\bk})}}
, \nonumber \\
u_{\bk-}&=\frac{A_{\bk}+\omega_{\bk}}{\sqrt{2\omega_{\bk}(A_{\bk}+\omega_{\bk})}},~
v_{\bk-}=\frac{B_{\bk}}{\sqrt{2\omega_{\bk}(A_{\bk}+\omega_{\bk})}}
.\end{align}
The $\gamma_+$ triplon, with spin along the
quantization axis, will be the focus of the HFP
treatment, since it interacts with no other species and lowers
its energy from the Zeeman splitting.
Finally, the constant part of the Hamiltonian is given by
\begin{align}
\epsilon_0 =
-\frac{3}{4}J_0\sbar^2 + \mu(1-\sbar^2)
-\frac{3}{2N_d}\sum_{\bk}
A_{\bk}
.\end{align}

In the limit of vanishing inter-dimer interactions (in this case,
all $J',J'',J'''\to0$), the saddle-point
solution gives $\sbar =1$ and $\mu = -\frac{3}{4}J_0$.
This limit serves as a good starting point in Ba$_3$Cr$_2$O$_8$,
where $J_0$ is much larger than
the inter-dimer couplings.
Furthermore,
with these values of $\sbar$ and $\mu$,
the triplon dispersion matches the RPA form
fitted to experimental values in Ref.\onlinecite{kofudispersion}.
Solving the saddle-point conditions
with the couplings in Table \ref{tab:anisotropicinteractions}
taken as bare values, we find
$\sbar = 0.992$ and $\mu = -0.775J_0$, showing a
high degree of dimerization.
The bare couplings will be slightly renormalized (compared to
the $\sbar =1$, $\mu = -3J_0/4$ case) as a result.
In particular, $J_0\to J_0-\Delta\mu$ and
$J_m\to\sbar^2J_m$ for $m\in\{1,\ldots,9\}$.
However, we can take these renormalized
couplings to have the values in Table \ref{tab:anisotropicinteractions},
since only the final dispersion will be used in
our Hartree-Fock calculation.
We furthermore assume that the dispersion is not temperature dependent
within the low-temperature regime considered.

\section{Hartree-Fock Effective Hamiltonian}
\label{sec:hfp}

We turn our focus to the $\gamma_{+}$ quasiparticle; being the
field-aligned quasiparticle, it will condense with sufficiently
large Zeeman splitting. We consider field strengths large enough that
we may ignore the higher-energy $\gamma_0$ and $\gamma_-$
quasiparticles. The typical splitting is of energy
$g\mu_bH_c(0) \cong$ 15.4 K.
This scale is significantly larger than the highest temperature (around 2.7 K)
where the BEC transition occurs\cite{aczeldata}.
Consequently, ignoring terms in the Hamiltonian
with $\gamma_0$ and $\gamma_-$
is a safe approximation to make
in this external field regime.

The Hamiltonian we take for the $b \equiv \gamma_+$ triplons is
$\HM = \HM_K + \HM_U$, which is the sum of the kinetic and inter-triplon
interaction terms. Here,
\begin{align}
&\HM_K = \sum_{\bk} \left( \epsilon_k - \mu \right) b^{\dag}_{\bk}b_{\bk},
\\
&\HM_U = \frac{1}{2N_d}\sum_{\bk,\bk',\bq}U_{\bq}
b^{\dag}_{\bk}b^{\dag}_{\bk'}b_{\bk+\bq}b_{\bk'-\bq},
\label{eqn:hkplushu}
\end{align}
where $\mu = g\mu_BH - \Delta$ is the chemical potential
and $\Delta$ is the zero-field gap (1.37 meV from the
bond-operator theory).
$\epsilon_{\bk} + \Delta$ is the zero-field dispersion
with $\epsilon_{\bk}$ determined from the bond-operator theory.
The quartic terms from the bond-operator theory give rise to interactions
between the $b$ triplons. Combined with the interaction
from the hard-core constraint, this gives an approximate form for
$U_{\bq}$.
However, in the low-temperature
limit where the excited triplons
lie near the band minimum at $\bQ$, we may approximate
$U_{\bq} \approx U_{\bQ}$ as a constant, $U$. The value of the interaction
parameter $U$ will be
determined from a fit to the experimental data.

The condensate will form at the dispersion minimum
$\bQ=\frac{1}{2}(\v{u}+\v{v})$ \cite{kofutransition}.
Note that we define the reciprocal lattice vectors $\v{u}$, $\v{v}$, and $\v{w}$ in the conventional way.
For example,
$\v{u} = 2\pi\frac{\v{b}\times\v{c}}
{\v{a}\cdot(\v{b}\times\v{c})}$.
We follow the Hartree-Fock-Popov approach of Ref. \onlinecite{misguich}
by condensing the triplons
at $\bQ$: $b^{\dag}_{\bQ},~b_{\bQ}\to\sqrt{N_dn_c}$, where
$n_c$ is the condensate density (the condensed boson fraction
per dimer).
Introducing the summation $\sum\!'$, which excludes any terms containing creation or annihilation
operators at momentum $\bk=\bQ$, we decompose $\HM_U$ as follows:
\begin{align}
\HM_U &= \frac{U}{2N_d}N_c^2 +
\frac{UN_c}{N_d}\sum_{\bq}\!'\left\{\frac{b_{\bq}b_{-\bq}
+ b^{\dag}_{-\bq}b^{\dag}_{\bq}}{2}+2b^{\dag}_{\bq}b_{\bq}\right\}
\nonumber \\
&+\frac{U\sqrt{N_c}}{N_d}\sum_{\bk,\bq}\!'\left\{b^{\dag}_{\bk}b_{\bk+\bq}
b_{\bQ-\bq} + h.c.\right\}  \nonumber \\
&+\frac{U}{2N_d}\sum_{\bk,\bk',\bq}\!'b^{\dag}_{\bk}b^{\dag}_{\bk'}b_{\bk+\bq}
b_{\bk'-\bq},
\end{align}
using the fact that $2\bQ$ is a reciprocal lattice vector, so that
$b_{-\bk+\bQ}=b_{-(\bk+\bQ)}$.
Performing a mean-field
quadratic decoupling of the quartic
terms yields the following mean-field quadratic Hamiltonian:
\begin{align}\label{eqn:meanfieldh}
\HM_{\textrm{MF}} = E_0
+\sum_{\bk}\!'\ekt b^{\dag}_{\bk}b_{\bk}
+\frac{Un_c}{2}\sum_{\bq}\!' \{b_{\bq}b_{-\bq}+b_{-\bq}^{\dag}b_{\bq}^{\dag}\},
\end{align}
where
\begin{align}\label{eqn:meanfieldconst}
&E_0 = -\mu n_c +  UN_d\left(
\frac{n_c^2}{2}-(n-n_c)^2\right),
\nonumber\\
&\ekt = \epsilon_{\bk}-\meff,
\nonumber\\
&\meff = g\mu_BH-\Delta - 2Un.
\end{align}
This decoupling is valid so long as the triplon densities
$\avg{b^{\dag}_{\bk}b_{\bk}}$ are small.
In the noncondensed (normal) phase, the Hamiltonian is
already diagonalized. The triplon density is
given by the Bose distribution function, and must be
determined self-consistently:
\begin{align}
n = \int\Dk f_B(\ekt).
\end{align}
In the condensed phase, we must perform another Bogoliubov
transformation, which leads to the following diagonalized Hamiltonian:
\begin{align}\label{eqn:quadcondensed}
\HM_{\textrm{MF}} = \sum_{\bk}\!'E_{\bk}\left(\varphi^{\dag}_{\bk}
\varphi_{\bk}\right) - \frac{1}{2}\sum_{\bk}\!'\ekt + E_0,
\end{align}
where
\begin{align}
&E_{\bk} = \sqrt{\ekt^2-(Un_c)^2},
\nonumber\\
&\varphi_{\bk}=\tu_{\bk}b_{\bk}+\tv_{\bk}b^{\dag}_{-\bk},
\end{align}
in which
\begin{align}
&\tu_{\bk}=\sqrt{\frac{\ekt}{2E_{\bk}}+\frac{1}{2}},\quad
\tv_{\bk}=\sqrt{\frac{\ekt}{2E_{\bk}}-\frac{1}{2}}.
\end{align}

The number of thermally excited triplons is
\begin{align}
n-n_c =
\int\Dk\left[\frac{\ekt}{E_{\bk}}\left(f_B(E_{\bk})+\frac{1}{2}\right)
\right]-\frac{1}{2}.
\label{eqn:ntilde}
\end{align}

For the final $\varphi$ quasiparticles to be condensed at $\bQ$, they
must be gapless. This constrains the effective chemical potential as
$\meff=-Un_c$, so that the field in the
condensed phase is given by
\begin{align}
g\mu_BH=\Delta+U\left(2n-n_c\right).
\label{eqn:hcondensed}
\end{align}

Between these two phases is the transition curve $H_c(T)$ where
triplons begin to condense.
The $b$ triplons are gapped in the noncondensed phase.
However, at the transition point to the condensed phase, they
become gapless. This constrains the effective chemical
potential as $\meff=0$ to give the critical field
\begin{align}
H_c(T) = \frac{\Delta}{g\mu_B} + \frac{2U}{g\mu_B}\ncr(T)
\label{eqn:hcrit}
.\end{align}
Since $\ekt = \epsilon_{\bk}$ in this case, we determine
the critical boson density at the transition, $\ncr$, by
the integral
\begin{align}
\ncr = \int\Dk f_B(\epsilon_{\bk})
\label{eqn:ncrit}
.\end{align}

\section{Critical Density Phase Diagram}
\label{sec:phasediagram}

In the HFP approach the critical field $H_c$ depends on the critical
boson density $\ncr$ linearly  as shown in
Eq.\eref{eqn:hcrit}.
A linear fit of $H_c$ to $\ncr$ determines the
interaction parameter $U$ from the slope.
In addition the zero-temperature critical field $H_c(0)$
gives another estimate for the gap $\Delta$.
With a given experimental data ($H_c,T$), we obtain
($H_c$,$\ncr$) pairs making use of the Eq.(\ref{eqn:ncrit}).
In the low-temperature, low-density regime where HFP approach is valid,
we expect $H_c$ to be linear in $\ncr$.

\begin{figure}[htp]
\includegraphics[width =  8 cm]{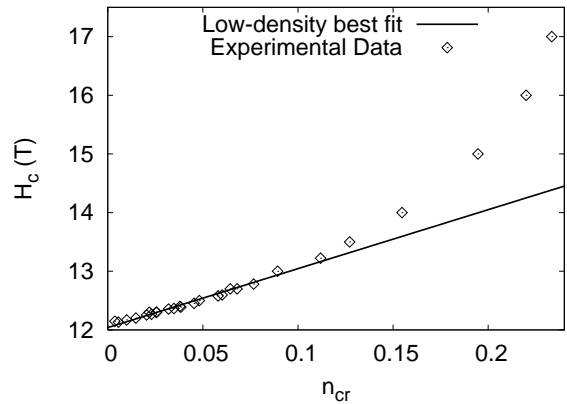}
\caption{
The critical field $H_c$
as a function of critical
density $\ncr$ with the applied field
$H$ parallel to the $c$-axis. $\ncr(T)$ is calculated, using the HFP approach, from
the given experimental temperatures.
A linear fit of $H_c$ to $\ncr$
is performed in a low-density range.
The experimental data is from M. Kofu \textit{et al.} in Ref.\onlinecite{kofutransition}
}
\label{fig:hcnckofuparr}
\end{figure}
\begin{figure}[htp]
\includegraphics[width =  8 cm]{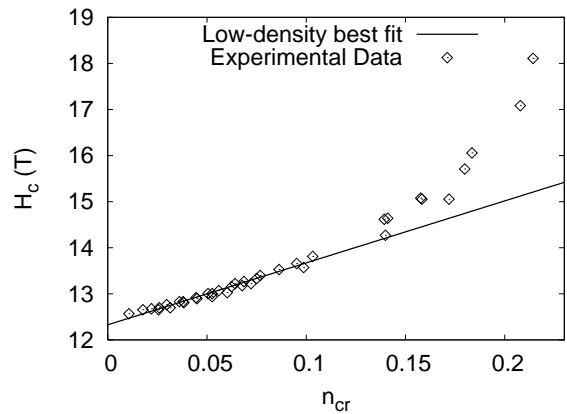}
\caption{
Same plot as in Fig.\ref{fig:hcnckofuparr}, but
the data is from A. A. Aczel \textit{et al.} in Ref.\onlinecite{aczeldata}.
}
\label{fig:hcncaczelparr}
\end{figure}
\begin{figure}[htp]
\includegraphics[width =  8 cm]{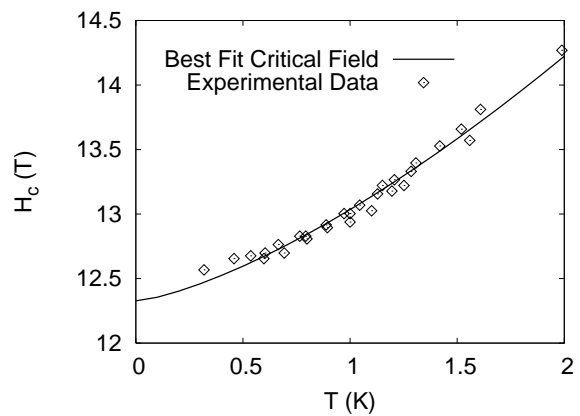}
\caption{Phase diagram giving the critical field $H_c$ as a
function of temperature. Experimental data is from
A. A. Aczel \textit{et al.} in Ref.\onlinecite{aczeldata},
with applied field $H$ parallel to
the $c$-axis.
The solid line shows the theoretical result obtained from the HFP approach
using the linear fit displayed in Fig.\ref{fig:hcncaczelparr}.
}
\label{fig:ncritaczel}
\end{figure}

We present three different fits of $H_c$ to $\ncr$.
These are based on the two different experimental data sets
obtained from
Ref.\onlinecite{kofutransition} and Ref.\onlinecite{aczeldata}.
Lines of best fit neglect the high-temperature
(density) regimes where $H_c$ loses linearity in $\ncr$.
A linear relation between $H_c$ and $\ncr$ is achieved for the case
of $H$ parallel to the $c$-axis, as shown in Fig.\ref{fig:hcnckofuparr}
and Fig.\ref{fig:hcncaczelparr}.
In Fig.\ref{fig:hcnckofuparr} we describe the result
of the linear fit which is obtained using the experimental data
from M. Kofu \textit{et al.} in Ref.\onlinecite{kofutransition}.
From the linear relation, we estimate the interaction constant
$U\cong6.5$ K and zero-field spin gap $\Delta\cong1.35$ meV.
The same analysis is performed using the second experimental data set
given by A. A. Aczel \textit{et al.} in
Ref.\onlinecite{aczeldata} and the result
is shown in Fig.\ref{fig:hcncaczelparr}.
This fit gives an estimate of $U \approx 8.7$ K and
$\Delta\cong1.34$ meV, which is very close to the value
obtained from the bond-operator approach.
Since this experimental data shows smaller
deviations from the linear fit, over a larger range
of temperatures,
we use the estimate of $U \approx 8.7$ K and
$\Delta\cong1.34$ meV in the following analysis.
In Fig.\ref{fig:ncritaczel} we display the low
temperature phase diagram $H_c(T)$, which is again obtained using
the data of Ref.\onlinecite{aczeldata}.

\begin{figure}
\includegraphics[width =  8 cm]{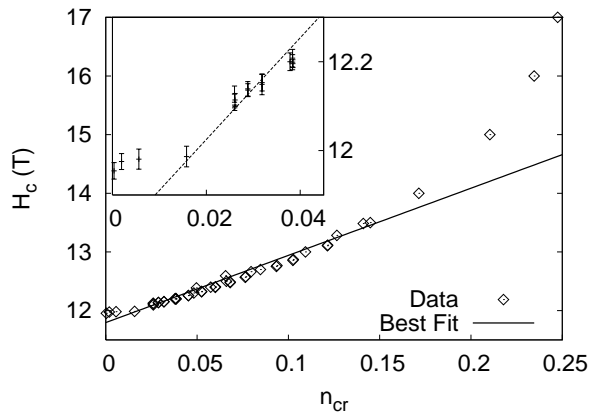}
\caption{
The critical field $H_c$ (circles) as a function of critical
density $\ncr$ with the applied field
$H$ perpendicular to the $c$-axis.
Here we use the data in Ref.\onlinecite{kofutransition} .
Inset: discrepancy between experiment and the best-fit line in
the small $\ncr$ regime.
}
\label{fig:perpfigs}
\end{figure}

On the other hand, the data for $H$ perpendicular to the $c$-axis
features low-temperature behaviour inconsistent
with the general linear trend as shown in Fig.\ref{fig:perpfigs}
and its inset.
We think that the existence of DM interaction is
one possible explanation of this low-temperature
discrepancy from the linear behavior.
Further discussion on this direction is shown in Sec.\ref{sec:discussion}.

\begin{figure}[htp]
\includegraphics[width =  8 cm]{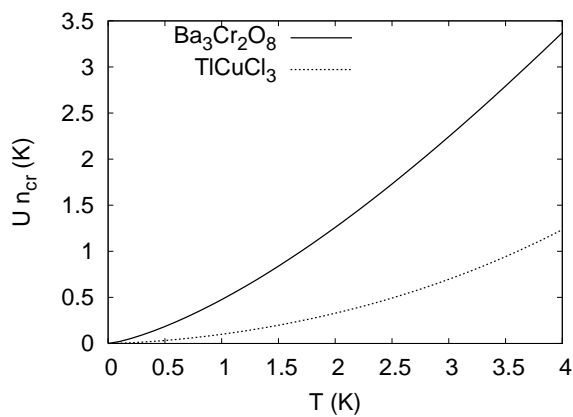}
\caption{Comparison of overall interaction energy scale
$U\ncr$ as a function of temperature
between
Ba$_3$Cr$_2$O$_8$ (solid line) and TlCuCl$_3$ (dashed line) systems.
$\ncr$ is the density of triplons at the condensate transition
and $U$ the inter-triplon interaction strength.
}
\label{fig:compuncr}
\end{figure}

Since a full-dispersion treatment successfully reproduces
the phase diagram for the dimerized spin system
TlCuCl$_3$,\cite{misguich}
it is instructive to use it in comparison with
the HFP approach for Ba$_3$Cr$_2$O$_8$.
Triplons in Ba$_3$Cr$_2$O$_8$ have a smaller self-interaction
constant, of $U\approx8.7$ K, compared to TlCuCl$_3$,
which has $U\approx 320$ K.\cite{misguich}
However, triplon densities are significantly higher in
Ba$_3$Cr$_2$O$_8$ than in TlCuCl$_3$, by over
an order of magnitude.
This makes
the Hartree-Fock critical field shift $U\ncr$
greater in Ba$_3$Cr$_2$O$_8$ than in TlCuCl$_3$.
In Fig.\ref{fig:compuncr} we plot
$U\ncr$ in these two systems with varying temperature.
Within the Hartree-Fock-Popov approach, the
term $-2Un$ acts as a shift in the effective chemical potential
as shown in Eq.(\ref{eqn:meanfieldconst}).
A decrease in $U$ will increase the
effective chemical potential, causing an increase in the
triplon density as seen with Ba$_3$Cr$_2$O$_8$.

\begin{figure}
\centering
\includegraphics[width =  8 cm]{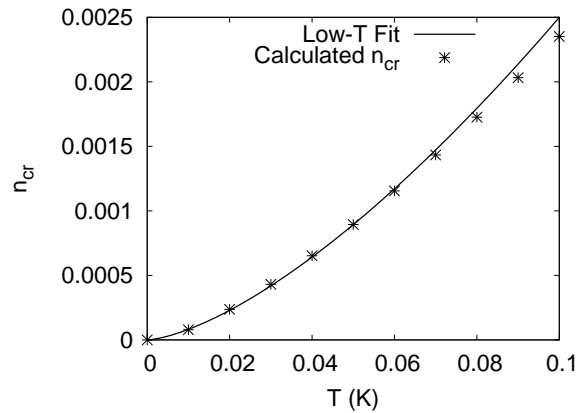}
\caption{Critical density $\ncr$ of Ba$_3$Cr$_2$O$_8$
calculated from the HFP theory with the full dispersion (cross points).
The solid line is $T^{\frac{3}{2}}$ power-law fit coming from the
simple quadratic dispersion.
The points obtained from the full dispersion begin
to deviate from the power-law fit around 0.06 K.
}\label{fig:bacroncrlogfits}
\end{figure}

\begin{figure}
\centering
\includegraphics[width = 8 cm]{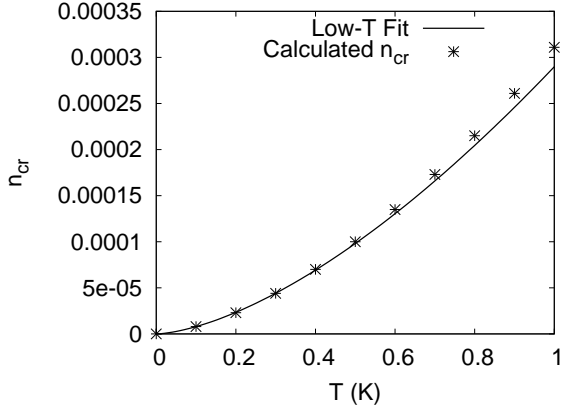}
\caption{Same plot as Fig.\ref{fig:bacroncrlogfits} but for the
TlCuCl$_3$ compound.
Note that the results obtained from the full dispersion
follow the power-law fit up to 0.6 K.
}
\label{fig:tlcuclncrlogfits}
\end{figure}

The shape of the dispersion affects the temperature range
in which the power-law behavior of
$\ncr\propto T^\frac{3}{2}$ is satisfied.
At very low temperature,
the quadratic approximation to the minimum of the dispersion becomes
very accurate. As $T\to0$, the quadratic dispersion
$\epsilon_{\bk} = \frac{\bk^2}{2m}$ yields
$\ncr\propto T^\frac{3}{2}$ by evaluating Eq.\eref{eqn:ncrit}
exactly\cite{nikuni} to give
\begin{align}
\lim_{T\to 0}\ncr(T)=\frac{\zeta_{\frac{3}{2}}}{2}
\left(\frac{Tm}{2\pi}\right)^{\frac{3}{2}}.
\label{eqn:ncrlowt}
\end{align}
In Ba$_3$Cr$_2$O$_8$, a low-temperature $T^\frac{3}{2}$ fit deviates from
the full disperion critical density
$\ncr$ around 0.06 K as shown in Fig.\ref{fig:bacroncrlogfits}.
This is an order of magnitude smaller than for TlCuCl$_3$,
where the $T^\frac{3}{2}$ behaviour persists up to
about 0.6 K as described in Fig.\ref{fig:tlcuclncrlogfits}.
The lower temperature scale of Ba$_3$Cr$_2$O$_8$ is expected to be
from the smaller triplon bandwidth, represented
by the large effective mass near the dispersion minimum.
From the power-law fits to Eq.\eref{eqn:ncrlowt}
in Fig.\ref{fig:bacroncrlogfits}
and Fig.\ref{fig:tlcuclncrlogfits}
we find that $1/m\cong1.36$ K (43.6 K) for Ba$_3$Cr$_2$O$_8$
(TlCuCl$_3$ \cite{misguich})
showing the narrower bandwidth of Ba$_3$Cr$_2$O$_8$.
Here we set $\hbar^2/k_B=1$.

\section{Specific Heat}
\label{sec:specheat}
We apply the HFP approach to explain the specific heat data measured
by M. Kofu \textit{et al}.\cite{kofutransition}
To determine the magnetic contribution to the specific heat,
we first find the expectation value of the
energy per dimer.
After condensing the triplons at momentum $\bQ$,
the diagonalized mean-field Hamiltonian
contains only number operators of
thermally distributed bosonic quasiparticles
(see Eqs. (\ref{eqn:meanfieldh}) and (\ref{eqn:quadcondensed})).
By differentiating the energy with respect to temperature,
we find the specific heat per dimer.
In the normal phase,
\begin{align}
\frac{\avg{E}}{N_d} = -Un^2 + \int\Dk\ekt f_B(\ekt)
\end{align}
and
\begin{align}
\frac{C_V}{N_dk_B} = -\beta\int\Dk\ekt^2\pd{f_B}{\ekt}
+2U\pd{n}{T}\int\Dk \ekt\pd{f_B}{\ekt},
\label{eqn:cvnormal}
\end{align}
with
\begin{align}
\pd{n}{T}=-\beta\frac{\int\Dk\ekt\pd{f_B}{\ekt}}{1-2U\int\Dk\pd{f_b}{\ekt}}.
\end{align}
However, in the condensed phase, we have
\begin{align}
\frac{\avg{E}}{N_d} = E_0-\frac{1}{2}\int\Dk\ekt +\int\Dk E_{\bk}
\left(f_B(E_{\bk})+\frac{1}{2}\right)
\end{align}
and
\begin{align}
&\frac{C_V}{N_dk_B} = -\beta\int\Dk E_{\bk}^2\pd{f_B}{E_{\bk}} +
2U\pd{n}{T} \times   \nonumber \\
&\left[
n_c-n-\frac{1}{2}+\int\Dk\frac{\epsilon_{\bk}}{E_{\bk}}\left(
f_B(E_{\bk})+\frac{1}{2}+E_{\bk}\pd{f_B}{E_{\bk}}\right)\right],
\label{eqn:cvcondensed}
\end{align}
with
\begin{align}
\pd{n}{T} = \frac{\beta\int\Dk\ekt\pd{f_B}{E_{\bk}}}
{1-2U\int\Dk\frac{\epsilon_{\bk}}{E_{\bk}^2}\left(
-\ekt\pd{f_B}{E_{\bk}}+ \frac{\meff}{E_{\bk}}
(f_B(E_{\bk})+\frac{1}{2}\right)}.
\end{align}

Non-magnetic contributions to the
specific heat, such as phonon contribution, will not change
appreciably with the applied field.
The difference $C_V(H)-C_V(0)$
thus captures the heat capacity contribution from triplons.
Currently, there exists no zero-field specific heat data,
preventing proper quantitative comparison.

\begin{figure}[htp]
\includegraphics[width =  8 cm]{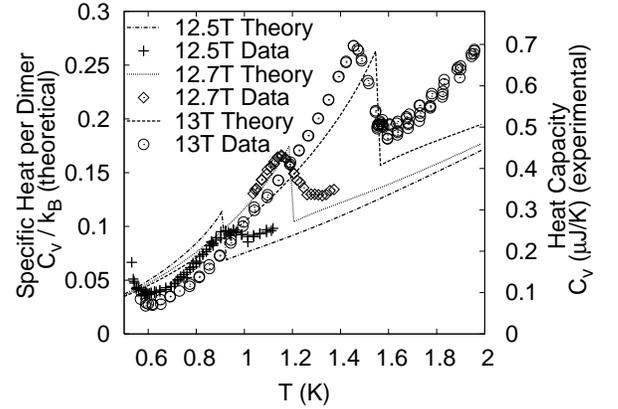}
\caption{
Comparison of experimental heat capacity to
theoretical specific heat per dimer
as a function of temperature. Comparisons are made for
external fields of 12.5 T, 12.7 T, and 13 T.
The experimental data is from Ref.\onlinecite{kofutransition} with the applied field
$H$ parallel to the $c$-axis.
}
\label{fig:cvkofuparr}
\end{figure}

However, we may still make a comparison, up to an overall scale
difference, between the theoretical specific heat and
experimental heat capacacity data.
Fig.\ref{fig:cvkofuparr}
shows the calculated magnetic contribution with the
experimentally determined heat capacity
in Ref.\onlinecite{kofutransition}. The relative scale is chosen to best
show similarity in the peak shape
for fields close to the zero-temperature critical field.
Despite the scale difference and non-triplon contribution,
the theoretical result still captures the peak at the critical
temperature. However, the drop in heat capacity is overestimated.
Furthermore, it is discontinuous, which can be considered as an artifact
of the HFP approximation.\cite{misguich}

\section{Magnetization}
\label{sec:magnetization}

When $H>H_c$, $\gamma_+$ bosons condense, leading to the macroscopic occupation
of the triplet states with the momentum
corresponding to the dispersion minimum. The ground state wave function
is then given by the coherent superposition of the
singlet and the $S_z$=1 triplet states.\cite{GiamarchiReview,matsumotoprl}
The density of the condensate
determines the magnetization along the z-direction.
In addition, the condensate supports the staggered
magnetization which has finite transverse components
$\langle {S_i}^{x}\rangle$ and $\langle {S_i}^{y}\rangle$
breaking the continuous U(1) rotation symmetry around
the z direction.

To determine the magnetic ordering, we begin by rewriting the
spin operators in terms of the $t_0$, $t_-$ and $t_+$ operators.
Using the bond operator representation we obtain the following relations:
\begin{align}
&(S_1+S_2)_{\alpha} = -i\epsilon_{\alpha\beta\gamma}t^{\dag}_{\beta}
t_{\gamma}, \nonumber \\
&(S_1-S_2)_{\alpha} = s^{\dag}t_{\alpha} + t^{\dag}_{\alpha}s.
\end{align}
for $\alpha\in\{x,y,z\}$.
Due to Zeeman splitting, the $t_z = \gamma_0$ triplets
are negligible, and we find that
$\avg{(S_1+S_2)_x} = \avg{(S_1+S_2)_y} = \avg{(S_1-S_2)_z} = 0$.
As $\gamma_-$ are similarly negligible, we  expand the rest of
the triplet $t_\pm$ operators in terms of the
$\gamma_\pm$ operators. After that we ignore
the terms with $\gamma_-$ because they do not contribute to expectation values.
The average spin component per dimer along the field direction,
which is nothing but the fraction of aligned quasiparticles $n$, is given by
\begin{align}
\avg{(\v{S}_1+\v{S}_2)_z} &= \avg{t^{\dag}_+t_+-t^{\dag}_-t_-}
\nonumber \\
&=\frac{1}{N_d}\sum_{\bk}\avg{\gamma^{\dag}_{\bk+}\gamma_{\bk+}}
(u_{\bk-}^2-v_{\bk-}^2) = n.
\end{align}
Since we have condensed singlet $\sbar$, the staggered component of
the spin becomes (using $u_{-\bk} = u_{\bk}$)
\begin{align}
\avg{S_{i1x}-S_{i2x}} &= \frac{\sbar}{\sqrt{2}}\avg{t_{i+}^{\dag}+
t_{i-}^{\dag}+t_{i+}+t_{i-}}  \nonumber \\
&=\frac{\sbar}{\sqrt{2N_d}}\sum_{\bk}e^{i\bk\cdot\v{r}_i}\avg{
t^{\dag}_{\bk+} + t^{\dag}_{\bk-}} + h.c.
\nonumber \\
&=\frac{\sgn(B)\sqrt{2}\sbar}{\sqrt{N_d}}(u_{\bQ-}-v_{\bQ-})
\Re(e^{i\bQ\cdot\v{r}_i}\Gamma_{\bQ})
.\end{align}

Here $\Gamma_{\bQ} = \avg{\gamma^{\dag}_{\bQ+}}$ with
$|\Gamma_{\bQ}|^2=n_c$.
Without loss of generality, we fix the overall
phase by taking $\Gamma_{\bQ}$ to be real.
Only the coherent condensate contributes to the
transverse magnetization.
Similarly, the $y$-component comes from the imaginary component of
the condensate,
\begin{align}
\avg{S_{i1y}-S_{i2y}} = \frac{\sgn(B_\bQ)\sqrt{2}\sbar}
{\sqrt{N_d}}(u_{\bQ-}-v_{\bQ-})\Im(e^{i\bQ\cdot\v{r}_i}\Gamma_{\bQ}).
\end{align}
The transverse spin component thus is spatially modulated by
the condensate wavevector $\bQ$.
The transverse magnetization per dimer can be written
as \cite{matsumotoprb}
\begin{align}\label{eqn:perpmagdefn}
M_{xy} &\equiv \frac{1}{N_d}\sum_ie^{i\bQ\cdot\v{r}_i}
\avg{S_{i1x}-S_{i2x}}
\\
&=\frac{\sgn(B_\bQ)\sqrt{2}\sbar}
{\sqrt{N_d}^3}\left(u_{\bQ-}-v_{\bQ-}\right)
\sum_ie^{i\bQ\cdot\v{r}_i}\cos({\bQ\cdot\v{r}_i})\Gamma_\bQ
\nonumber \\
&=\frac{\sgn(B_\bQ)\sbar}{\sqrt{2N_d}}\left(u_{\bQ-}-v_{\bQ-}\right)
\Gamma_\bQ.
\end{align}
The square of the transverse magnetization per Cr$^{5+}$ ion is then
\begin{align}
M^2_{\perp}&=\left(g\mu_BM_{xy}\right)^2= g^2\mu_B^2
\frac{\sbar^2\Gamma_\bQ^2}{8N_d}
\frac{(\omega_{\bQ}+A_{\bQ}-B_{\bQ})^2}
{2\omega_{\bQ}(A_{\bQ}+\omega_{\bQ})}
\nonumber \\
&=g^2\mu_B^2
\frac{\sbar^2n_c}{4}\frac{A_{\bQ}-B_{\bQ}}{2\omega_{\bQ}}
=
\sbar^2n_c\frac{J_0g^2}{8\Delta}\mu_B^2
.\end{align}
Having neglected the $\gamma_0$ and $\gamma_-$
triplons, we estimate
$\sbar^2\cong 1-n$, using the overall triplet boson constraint.
The total and condensed triplet densities,
$n$ and $n_c$, are determined by solving Eq.\eref{eqn:ntilde}
and Eq.\eref{eqn:hcondensed} self-consistently.

\begin{figure}[htp]
\includegraphics[width =  8 cm]{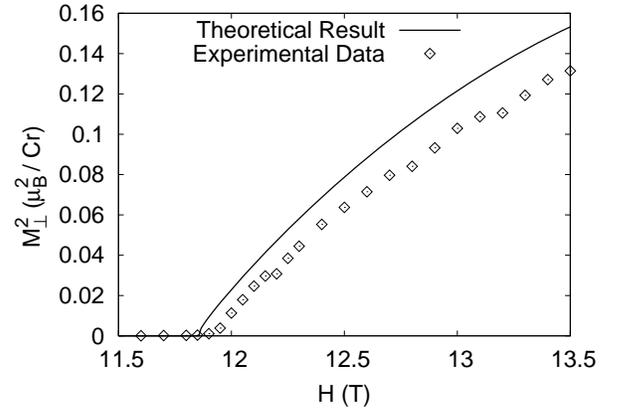}
\caption{Comparison of experimental and theoretical
perpendicular magnetization squared at $T = 0.2$ K.
Perpendicular magnetization is defined in Eq.\eref{eqn:perpmagdefn}.
The experimental data is from Ref.\onlinecite{kofutransition}, with the applied field
$H$ perpendicular to the $c$-axis.}
\label{fig:mp0.2}
\end{figure}

\begin{figure}[htp]
\includegraphics[width =  8 cm]{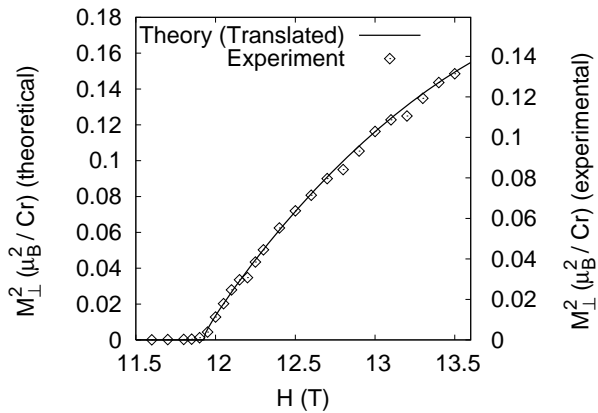}
\caption{Comparison of the shape of theoretical and experimental
perpendicular magnetization curves as in Fig. \ref{fig:mp0.2}.
Theoretical result has been
translated by $0.07$ T to match the experimental critical-field
behaviour. The scale of the theoretical result is 1.13 larger than
that of the experimental data.}
\label{fig:scalemperp0.2}
\end{figure}

The transverse magnetization has been measured
by the elastic neutron scattering experiments.\cite{kofutransition}
The applied field is perpendicular
to the $c$-axis.
Fig.\ref{fig:mp0.2} compares theoretical squared
perpendicular magnetization
at $T=0.2$ K to the experimental results.\cite{kofutransition}
Deviation
from the experiment occurs most prominently in the
critical field. This is caused by discrepancy
between the linear fit
$H_c\propto\ncr(T)$ and the experimental critical field $H_c(T)$.
However, the shape of the magnetization curve past the critical
field is properly reproduced. This can be seen in Fig.\ref{fig:scalemperp0.2},
where the theoretical result has been
translated to match the experimental critical field.
The resulting shape matches over the entire range of fields, with the
theoretical magnetization larger by
a factor of 1.13.
The magnetization also jumps slightly
at the critical field. Like the discontinuity in specific heat,
this is an artifact of
the HFP treatment.\cite{sirker05}

\begin{figure}[htp]
\includegraphics[width =  8 cm]{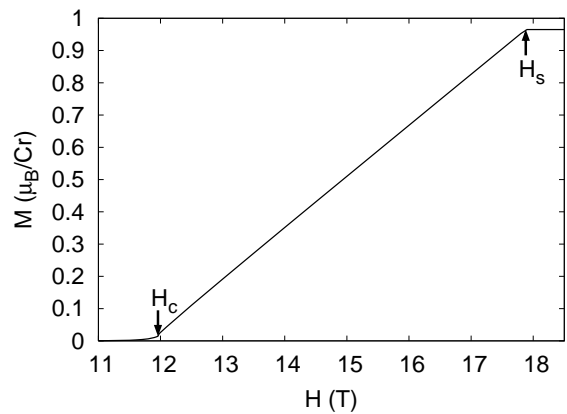}
\caption{Parallel magnetization per Cr atom, as a function
of applied field perpendicular to the $c$-axis, at $T=0.4$ K.
The result is found from the triplon density as determined
by the HFP approach. Critical field $H_c$ and saturation field
$H_s$ are indicated.}
\label{fig:mparr}
\end{figure}

The parallel magnetization has been measured as a function
of applied field (both parallel and perpendicular to the
$c$-axis) at the condensate transition.\cite{kofutransition}
Fig.\ref{fig:mparr}
gives the HFP result for $H$ perpendicular to the $c$-axis,
with the magnetization per dimer
$M_\parallel =\frac{g}{2}\mu_Bn$.
The saturation field, where all spins are aligned with the field,
is severely underestimated by the HFP result
of 18 T. Experimentally it is found between 23 T and 24 T,
from the derivative of magnetization $\pd{M}{H}$.
\cite{kofutransition}
This happens because the triplon density in the HFP
approach grows too quickly with increasing field.

\section{Discussion}
\label{sec:discussion}

We have applied the HFP approach to understand the triplon
BEC in Ba$_3$Cr$_2$O$_8$, using the full dispersion of the
triplons measured in the recent neutron scattering experiments
(which is recast in the form of a bond-operator
representation of a Heisenberg model).
We investigated the temperature range where the HFP approach
is valid with the full dispersion, and also locates the temperature
where the quadratic approximation of the dispersion breaks down.
Using this approach, we computed the transverse magnetization
and specific heat that are favorably compared to available
experimental data. Our results show that the BEC picture overall
works reasonably well for Ba$_3$Cr$_2$O$_8$.

In the much-studied three-dimensionally-coupled spin-dimer system
TlCuCl$_3$, the triplon band width $W \sim 87$ K and the effective
interaction $U \sim 340$ K within the HFP analysis.\cite{misguich}
In contrast, our analysis leads to
$W \sim 21$ K and $U \sim 8.7$ K in Ba$_3$Cr$_2$O$_8$.
Thus it may appear that the HFP would work better for
Ba$_3$Cr$_2$O$_8$ because of smaller $U/W$.
On the other hand, smaller $U$ in Ba$_3$Cr$_2$O$_8$ results in
a larger critical triplon density $n_{cr} \sim 0.1$ compared to
$n_{cr} \sim 0.002$ in TlCuCl$_3$, making the
dilute triplon density approximation less valid.
In the end, the combined effect in the form of the
HFP correction to
the critical field $H_c(T)$, $U n_{cr}$, turns out to be bigger for
the case of Ba$_3$Cr$_2$O$_8$.
This means that the temperature range where the HFP
approach is valid is more limited in the case of Ba$_3$Cr$_2$O$_8$.
Indeed, it is found that the HFP works up to $8K$ in TlCuCl$_3$
while it fits the data up to $2K$ at best in Ba$_3$Cr$_2$O$_8$.

The triplon dispersion in Ba$_3$Cr$_2$O$_8$ is flatter (or the effective
mass is larger) compared to TlCuCl$_3$, which leads to
a smaller window of temperatures where the quadratic-dispersion
approximation is valid.
This is seen in how the relation
$[H_c(T)-H_c(0)] \propto T^{3/2}$
reproduces the phase diagram for $T < 0.1$ K for
Ba$_3$Cr$_2$O$_8$, but for $T < 1$ K in TlCuCl$_3$.

A useful way to improve the HFP results may be to introduce the
hard-core constraint among the triplons.
The so-called Bruckner bond operator
approach\cite{roschvojta,kotov98} achieves this
by introducing an infinite on-site triplon repulsion by
\begin{align}
\HM_{C} = V\sum_{i\alpha\beta}
t^{\dag}_{i\alpha}t^{\dag}_{i\beta}t_{i\alpha}t_{i\beta}
\end{align}
as $V\to\infty$.
In the low-density limit, this hard-core interaction may be
treated exactly by a summation of ladder diagrams
at the one-loop level in the self-energy.
This approach, when generalized to finite temperature,
should lower the triplon densities and extend
the region where low-density approximations are valid.
This may be a useful future extension of our work.

Recent ESR measurements indicate the existence of siglet-triplet
mixing in the ground state of Ba$_3$Cr$_2$O$_8$.
\cite{kofutransition}
In the ground state, singlets mix with $t_0$ for $H\perp c$,
and with $t_\pm$ for $H\parallel c$.
This mixing points to the existence of a Dzyaloshinsky-Moriya (DM)
interaction of the form $\v{D}_{ij}\cdot\v{S}_i\times\v{S}_j$, with
$\v{D}_{ij}$ perpendicular to the $c$-axis.
Since it breaks the U(1)
symmetry of the Heisenberg Hamiltonian, the system is no longer
described by a BEC transition. The result is that triplons are
gapped and always condensed to some extent, turning the
transition into a crossover region.\cite{sirker04,sirker05}
Below the temperature scale of the DM interaction, then,
we expect that a simple BEC picture of triplons will
no longer be sufficient.
This could explain, for instance, the nonlinearity of
critical field $H_c$ in critical density $\ncr$ at low
temperatures for $H$ perpendicular to the $c$-axis.
An understanding of the magnitude and direction of the
$\v{D}_{ij}$ vector is important for a proper and full description of
Ba$_3$Cr$_2$O$_8$, especially
at very low temperatures.

\acknowledgments

We thank S. H. Lee for providing the experimental data on Ba$_3$Cr$_2$O$_8$ and
many helpful discussions.
This work was supported by the NSERC of Canada, the Canada Research Chair program,
and the Canadian Institute for Advanced Research.
We also acknowledge the hospitality of the Kavli Institute for Theoretical Physics and
the Aspen Center for Physics, where various parts of this work were performed.



\end{document}